\documentclass[twoside]{ilcws07}
\usepackage[latin1]{inputenc}
\usepackage[dvips]{graphicx,epsfig,color}
\usepackage{wrapfig,rotating}
\usepackage{amssymb,amsmath,array}
\usepackage{cite}

\begin{document}
\title{Tools for NNLO QCD Calculations} 
\author{T.\ Gehrmann
\vspace{.3cm}\\
Institut f\"ur Theoretische Physik,
Universit\"at Z\"urich, CH-8057 Z\"urich, Switzerland
}

\maketitle

\begin{abstract}
For precision studies with QCD observables at colliders, higher order 
perturbative corrections are often mandatory. For exclusive observables, 
like jet cross sections or differential distributions, these corrections 
were until recently only known to next-to-leading order (NLO) in 
perturbation theory. Owing to many new technical developments, first 
next-to-next-to-leading order (NNLO) QCD calculations are now becoming 
available. We review the recent progress in this field. 
\end{abstract}

\vspace{-7.6cm}
{\noindent ZU-TH 22/07}
\vspace{7cm}
\section{Introduction}
At present-day colliders, a number of benchmark reactions is measured to 
a very high experimental accuracy. These reactions allow for a very precise
determination of the parameters of the Standard Model, and may reveal minute 
deviations, hinting towards new physics effects. 
For many of these benchmark reactions, one faces the problem that the
observables~\cite{dissertori}
 are inherently exclusive (like jet cross sections), 
or differential in several variables (if kinematical cuts are involved). 
While fully inclusive quantities are often known to third order in
perturbation theory, the perturbative expression for 
those  exclusive observables is much less well known. 
Consequently, 
the precision of many benchmark reactions  is limited not by the
quality of the experimental data, but
by the error on the theoretical (next-to-leading order, NLO)
calculations used for the extraction of
the Standard Model parameters. To
improve upon this situation, an extension of the theoretical calculations to
next-to-next-to-leading order (NNLO) is therefore mandatory.

\section{Ingredients to NNLO calculations}

At next-to-next-to-leading order, three types of processes contribute 
to an observable with $m$ hard particles in the final state:
\begin{enumerate}
\item the two-loop corrections to the $m$-particle process, where all 
particles are hard.
\item the one-loop corrections to the $(m+1)$-particle process, where one 
particle can become unresolved (collinear or soft).
\item the tree-level $(m+2)$-particle process, where up to two particles 
can become unresolved. 
\end{enumerate}
Each contribution contains infrared singularities, which cancel only in the 
sum. 
Since the definition of the observable final state (jet algorithm or
kinematical cuts) acts differently on each of these processes, it is 
mandatory to compute all of them individually. This is in variance with
calculations of fully inclusive quantities (total cross sections, sum rules),
where all contributions can be added before evaluation 
using the optical theorem. 

QCD infrared factorisation predicts that the behaviour of 
cross sections in soft or collinear limits is process-independent. At NNLO,
the corresponding universal tree-level double unresolved~\cite{campbell} 
and one-loop single unresolved~\cite{onelstr} factors are known. Based on this 
universal behaviour, one aims to develop process-independent techniques 
for NNLO calculations of exclusive observables. 

The first calculations combining elements of this type are the derivations
of the NNLO corrections to the deep inelastic structure functions~\cite{zij},
and to the coefficient functions for vector boson~\cite{mat} and Higgs 
boson production~\cite{higgs,pstricks} at hadron colliders. Although these 
observables are fully inclusive, one  still has to consider the individual 
parton-level contributions separately, since they include collinear radiation 
associated with the incoming partons. 

NNLO calculations for hadron colliders also require parton distributions 
accurate to this order. Following the derivation of the three-loop 
splitting functions~\cite{mvv}, global NNLO fits~\cite{mrst} are now becoming 
available. These are however still somewhat limited by the fact that not all
observables included in the fit are known to NNLO.

\section{Techniques and applications}
The derivation of the  individual ingredients to NNLO calculations, 
and their combination into a parton-level event generator, are posing 
a variety of computational challenges. Many new techniques have been 
developed in this context. 

\subsection{Sector decomposition}
In computing the various contributions to perturbative corrections at NNLO and 
beyond, one frequently encounters the problem of overlapping singularities. 
These appear if one particular term develops a singular behaviour in 
more than one region of phase space.
The technique of sector decomposition offers 
an elegant solution to this problem. 
Originally proposed as a tool in formal proofs of 
renormalisability~\cite{hepp},  and only used occasionally 
afterwards~\cite{roth}, this technique was fully formulated first for 
multi-loop integrals~\cite{secdecvirt} and subsequently extended to 
phase space integrations~\cite{secdecreal,ggh}.  

Starting from a parameter representation of the (virtual or real)
integral under consideration, 
 the space of 
integrations is decomposed iteratively into non-overlapping sectors. 
At the end of this iteration
each sector contains only a single type of singularity. 
In each sector, after expanding all regulators in distributions, 
the Laurent series of the integral is well-defined, and its 
coefficients can be obtained by numerical integration. 

This technique has been used in the calculation of virtual two-loop and 
three-loop multi-leg integrals~\cite{secdecvirt,studerus}, 
often ahead of or in 
timely coincidence with analytical results. For loop integrals involving 
mass thresholds inside the physical region, sector decomposition must be 
combined with contour deformation to avoid pinch 
singularities~\cite{petriello,beerli}. 

Concerning the calculation 
of real radiation corrections at NNLO, the first-ever results for fully
differential observables were obtained using sector decomposition for 
$e^+e^- \to 2$~jets~\cite{babis2j},
$pp\to H+X$~\cite{babishiggs}, muon decay~\cite{babismu} and
$pp\to V+X$~\cite{babisdy}.

\subsection{Integral reduction}
Within dimensional
regularisation,  the large number
of different integrals appearing in  
multi-loop calculations can be reduced
to a small number of  so-called {\em master integrals} by using
integration-by-parts (IBP)
identities~\cite{chet}. These identities
exploit the fact that the integral over the total derivative of 
any  of the loop
momenta vanishes in dimensional regularisation.

For integrals involving more than two external legs,  another
class of identities exists due to  Lorentz invariance.
 These
Lorentz invariance identities~\cite{gr} rely on the fact that  an
infinitesimal Lorentz transformation commutes with the loop integrations, thus
relating different integrals. The common origin of IBP and LI 
identities is the Poincare invariance of loop integrals within
dimensional regularisation.

In principle, the identities for all  loop integrals of a given topology can 
be solved in a closed symbolic form~\cite{chet}. In practise, this solution is
often overly complex, and can not be found in an automated manner. 
Instead, one pursues another approach, the so-called Laporta 
algorithm~\cite{laporta},
which solves the system of identities for a given topology by assigning
a weight to each integral according to its complexity. In turn, the 
very large system of interconnected identities is solved by elimination,
thereby reducing each integral in the system to master integrals.  
Several computer implementations of the Laporta algorithm are 
available~\cite{gr,laporta,air}. 

Originally formulated only for loop integrals, these reduction techniques 
can also be applied to multi-particle phase space integrals by expressing 
on-shell conditions and kinematic constraints in the form of 
generalised propagators~\cite{pstricks,ggh}. 

\subsection{Mellin-Barnes integration}
The techniques described in the previous section allow to reduce the 
hundreds to thousands of different integrals appearing in an actual 
calculation to a small set of master integrals. The
reduction equations do not yield any information
on the master integrals, which have to be determined using some other 
technique.
Unlike in the one-loop case, where a direct integration using Feynman 
parameters is usually sufficient, master integrals at two or more 
loops pose a considerable computational challenge. 

A commonly used analytical approach is to replace the product of loop 
propagators by multiple Mellin-Barnes integrations~\cite{smirnovbook}. 
This replacement can be performed in an automated manner using computer 
algebra~\cite{ambre}. After integration over the loop momenta in 
dimensional regularisation, the Mellin-Barnes integrals are normally 
not yet well-defined around $d=4$, where the Laurent expansion is desired. 
To arrive at this limit, one must first perform analytic continuations, 
which can again be done using computer algebra~\cite{daleo,mb}, thereby 
transforming each single Mellin-Barnes integral into a sum of integrals and
residues. After analytic continuation, the Laurent coefficients 
can be determined by carrying out the Mellin-Barnes integrals 
numerically~\cite{daleo}. Analytical results for Mellin-Barnes integrals 
can often be obtained (especially if they can be expressed as 
multiple
harmonic sums~\cite{nested,xsummer,blumlein}), 
and systematic methods for them are under 
development.

Using Mellin-Barnes integration techniques, analytical results were obtained 
for two-loop four-point functions with 
massless~\cite{smirbox,tausk,oleari2} 
and massive~\cite{gluza} internal propagators. 

\subsection{Differential equations}
A method for the analytic
computation of master integrals avoiding the explicit
integration over the loop momenta is to derive differential equations in 
internal propagator masses or in external momenta for the master integral, 
and to solve these with appropriate boundary conditions. 
This method has first been suggested~\cite{kotikov} to relate 
loop integrals with internal masses to massless loop integrals. 

It has been worked out detail and generalised to differential 
equations in external momenta in~\cite{gr,remiddi}. Differentiation of 
a master integral with respect to an external invariant or to a mass yields a
combination of integrals of the same topology. Using the integral reduction 
techniques described above, these can be reexpressed by the master integral 
itself, plus simpler integrals. As a result, one obtains an inhomogeneous 
differential equation for the master integral. Solving this equation 
and matching the solution onto an appropriate boundary condition (obtained 
in a special kinematical point) then yields the desired master integral,
very often in a closed form containing hypergeometric functions and their 
generalisations. The Laurent expansion of the integral then amounts 
to expansion of these functions~\cite{hypexp}.
A detailed review of the method can be found in~\cite{mastrolia}.

Using the differential equation technique, master integrals were 
derived for 
massless two-loop four-point functions~\cite{oleari1,bastei,3jmi},
for massive two-loop vertex functions~\cite{bonciani,kunszt} and for 
parts of the master integrals required for Bhabha scattering~\cite{bhabhade}.

\subsection{Virtual corrections and infrared structure}
Using the IBP and LI equations for the reduction to master integrals, which
were then computed with various of the above-mentioned techniques, results for 
a variety of two-loop corrections were obtained for $1\to 2$, $1\to 3$
and $2\to 2$ reactions. These include all massless parton-parton scattering 
amplitudes~\cite{m1}, processes yielding two-photon final states~\cite{m2},
light-by-light scattering~\cite{m3} 
and vector boson decay into three massless partons~\cite{3jme,muw} and its 
crossings~\cite{ancont}. For amplitudes involving external masses, all
two-loop form factors~\cite{breuther} of a heavy quark were derived, and 
large parts of the virtual 
two-loop corrections to Bhabha scattering~\cite{bhabha} are completed.  

These results are usually expressed in terms of 
harmonic polylogarithms~\cite{hpl}, which are a generalisation of 
the well-established Nielsen's polylogarithms~\cite{nielsen}. 
If several scales are involved 
in a process, the set of functions needs to be further extended to 
include multi-dimensional harmonic
polylogarithms~\cite{3jme,tdhpl,weinzierl}.

After ultraviolet renormalisation, these two-loop amplitudes still contain 
poles of infrared origin. This infrared pole structure is 
universal for massless amplitudes, and can 
be predicted from resummation formulae~\cite{catani}. Exploiting the fact that 
a particle mass can also act as an infrared regulator, the universality 
of infrared singularities can be extended to massive amplitudes~\cite{mm},
where not only the divergent terms but also logarithmically enhanced terms  
are universal. Knowing the corresponding massless scattering amplitudes, it 
is therefore possible to construct massive amplitudes up to corrections of 
order $m^2/s$, which was recently accomplished for heavy quark 
production~\cite{top} and Bhabha scattering~\cite{penin,becher}.  

\subsection{Subtraction methods}
To build exclusive final states at a given order, a jet algorithm or 
event shape definition 
has to
be applied separately to each partonic channel contributing at this order
and all partonic channels have to be summed.
However, each partonic channel contains infrared singularities which,
after summation, cancel among each other.
Consequently, these infrared singularities have to be extracted before
the jet algorithm can be applied.
While explicit infrared singularities from purely virtual
contributions are obtained immediately after integration over the loop
momenta, their extraction is more involved for real radiation.
The singularities associated with the real emission of soft and/or collinear
partons in the final state become only explicit after integrating
the real radiation matrix elements over the appropriate phase space.

In the sector decomposition method described above, the full matrix elements 
are integrated numerically, after expansion as a Laurent series
(which amounts to subtraction of residues in each sector). A different 
approach is pursued by subtraction methods, which 
extract 
 infrared singularities of the real radiation contributions 
 using infrared subtraction terms.
These terms are constructed such that they approximate the
full real radiation matrix elements in all singular limits while still
being integrable analytically.

Several methods for  constructing
 NLO subtraction terms systematically  were proposed in the 
literature~\cite{sub,cs,singleun,ant,hadant}. For some of these methods, 
extension to NNLO was discussed~\cite{nnlosub} 
and worked out for special cases~\cite{cghiggs}. 
Up to now, the only method worked out in full detail to NNLO is antenna
subtraction~\cite{ourant}.

The basic idea of the antenna subtraction approach is to construct 
the subtraction terms  from antenna functions. 
Each antenna function encapsulates 
all singular limits due to the 
 emission of unresolved partons between two colour-connected hard
partons. All antenna functions can be derived systematically from matrix 
elements~\cite{our2j} for physical processes.
The antenna subtraction method was used recently in the derivation of 
the NNLO QCD corrections to $e^+e^- \to 3$~jets and related event 
shapes~\cite{ourT}.

\section{Results and outlook}
The computational challenges of NNLO QCD calculations required for precision 
phenomenology have led to the development of a variety of new technical 
methods. Using these methods, many ingredients to NNLO QCD calculations 
were assembled in recent times. Many core results were derived more than 
once using independent, different methods. 

These ingredients are now assembled into complete NNLO QCD calculations, which 
are usually carried out in the form of a parton-level event generator. In such 
a program, the full kinematical information is available for each event, 
and cross sections are obtained by adding all events relevant to the
observable under consideration. This setup allows a great flexibility in 
the implementation of experimental cuts, and  in the simultaneous evaluation 
of numerous kinematical distributions.

Up to now, fully exclusive NNLO QCD calculations were performed 
for  $e^+e^- \to 2j$~\cite{babis2j,our2j,weinzierl2j},  
$e^+e^- \to 3j$~\cite{ourT,our3j}, 
the forward-backward asymmetry in $e^+e^-$ annihilation~\cite{afbnnlo},
Higgs production~\cite{babishiggs,cghiggs}    
and vector boson production~\cite{babisdy} at hadron colliders, 
as well as muon 
decay in QED~\cite{babismu}.
 
Virtual two-loop corrections are available for a number of further $2\to 2$ 
reactions, including jet production 
and vector-boson-plus-jet production at hadron colliders, jet production 
in deep inelastic scattering, and heavy quark production observables. 
The full calculation of these observables may require further improvements 
to existing tools, but further results of NNLO calculations are 
 clearly within reach in the near future.

\section*{Acknowledgement}
This research was supported in part by the Swiss National Science Foundation
(SNF) under contract 200020-109162.

\begin{footnotesize}

\end{footnotesize}


\begin{thebibliography}{99}

\bibitem{dissertori}
R.K.\ Ellis, W.J.\ Stirling and B.R.\ Webber, {\it QCD and Collider Physics},
Cambridge University Press (Cambridge, 1996);\\
G.\ Dissertori, I.G.\ Knowles and M.\ Schmelling, {\it Quantum 
Chromodynamics: High Energy Experiments and Theory}, Oxford University Press
(Oxford, 2003).

\bibitem{campbell}
A.~Gehrmann-De Ridder and E.W.N.~Glover, Nucl.~Phys.\ B {\bf 517} (1998) 269;\\
J.\ Campbell and E.W.N.\ Glover,
Nucl.\ Phys.\ B {\bf 527} (1998) 264;\\
S.\ Catani and M.\ Grazzini, Phys.\ Lett.\ B {\bf 446} (1999) 143;
Nucl.\ Phys.\ B {\bf 570} (2000) 287.


\bibitem{onelstr}
Z.\ Bern, L.J.\ Dixon, D.C.\ Dunbar and D.A.\ Kosower,
Nucl.\ Phys.\ B {\bf 425} (1994) 217;\\
D.A.\ Kosower, Nucl.\ Phys.\ B {\bf 552} (1999) 319;\\
D.A.~Kosower and P.~Uwer, Nucl.\ Phys.\ B {\bf 563} (1999) 477;\\
Z.\ Bern, V.\ Del Duca and C.R.\ Schmidt, Phys.\ Lett.\ B {\bf 445}
(1998) 168;\\
Z.\ Bern, V.\ Del Duca, W.B.\ Kilgore and C.R.\ Schmidt, Phys.\ Rev.\ D
{\bf 60} (1999) 116001.


\bibitem{zij}
 E.B.~Zijlstra and W.L.~van Neerven,
  Nucl.\ Phys.\  B {\bf 383} (1992) 525;
  Phys.\ Lett.\  B {\bf 297} (1992) 377.


\bibitem{mat}
  R.~Hamberg, W.L.~van Neerven and T.~Matsuura,
  Nucl.\ Phys.\  B {\bf 359} (1991) 343;
{\bf 644} (2002) 403 (E).

\bibitem{higgs}
  R.V.~Harlander and W.B.~Kilgore,
  Phys.\ Rev.\ Lett.\  {\bf 88} (2002) 201801;\\
  S.~Catani, D.~de Florian and M.~Grazzini,
  JHEP {\bf 0105} (2001) 025;\\
  V.~Ravindran, J.~Smith and W.L.~van Neerven,
  Nucl.\ Phys.\  B {\bf 665} (2003) 325;\\
  O.~Brein, A.~Djouadi and R.~Harlander,
  Phys.\ Lett.\  B {\bf 579} (2004) 149;\\
  R.V.~Harlander and W.B.~Kilgore,
  Phys.\ Rev.\  D {\bf 68} (2003) 013001.

\bibitem{pstricks}
 C.~Anastasiou and K.~Melnikov,
  Nucl.\ Phys.\  B {\bf 646} (2002) 220;\\
C.~Anastasiou, L.J.~Dixon, K.~Melnikov and F.~Petriello,
  Phys.\ Rev.\  D {\bf 69} (2004) 094008.

\bibitem{mvv}
S.~Moch, J.A.M.~Vermaseren and A.~Vogt,
Nucl.\ Phys.\ B {\bf 688} (2004) 101;\\
A.~Vogt, S.~Moch and J.A.M.~Vermaseren,
Nucl.\ Phys.\ B {\bf 691} (2004) 129.

\bibitem{mrst}
  S.~Alekhin, K.~Melnikov and F.~Petriello,
  Phys.\ Rev.\  D {\bf 74} (2006) 054033;\\
 A.D.~Martin, W.J.~Stirling, R.S.~Thorne and G.~Watt,
  arXiv:0706.0459 [hep-ph].


\bibitem{hepp}
K.~Hepp,
Commun.\ Math.\ Phys.\  {\bf 2} (1966) 301.

\bibitem{roth}
M.~Roth and A.~Denner,
Nucl.\ Phys.\ B {\bf 479} (1996) 495.

\bibitem{secdecvirt}
T.~Binoth and G.~Heinrich,
Nucl.\ Phys.\ B {\bf 585} (2000) 741;
{\bf 680} (2004) 375.



\bibitem{secdecreal}
G.~Heinrich,
Nucl.\ Phys.\ Proc.\ Suppl.\  {\bf 116} (2003) 368;
 {\bf 135} (2004) 290;
  Eur.\ Phys.\ J.\  C {\bf 48} (2006) 25;\\
C.~Anastasiou, K.~Melnikov and F.~Petriello,
Phys.\ Rev.\ D {\bf 69} (2004) 076010;\\
T.~Binoth and G.~Heinrich,
Nucl.\ Phys.\ B {\bf 693} (2004) 134.


\bibitem{ggh}
A.~Gehrmann-De Ridder, T.~Gehrmann and G.~Heinrich,
Nucl.\ Phys.\ B {\bf 682} (2004) 265.

\bibitem{studerus}
 T.~Gehrmann, G.~Heinrich, T.~Huber and C.~Studerus,
  Phys.\ Lett.\  B {\bf 640} (2006) 252.


\bibitem{petriello}
 A.~Lazopoulos, K.~Melnikov and F.~Petriello,
  Phys.\ Rev.\  D {\bf 76} (2007) 014001.

\bibitem{beerli}
  C.~Anastasiou, S.~Beerli and A.~Daleo,
  JHEP {\bf 0705} (2007) 071.


\bibitem{babis2j}
 C.~Anastasiou, K.~Melnikov and F.~Petriello,
  Phys.\ Rev.\ Lett.\  {\bf 93} (2004) 032002.

\bibitem{babishiggs}
C.~Anastasiou, K.~Melnikov and F.~Petriello,
Phys. Rev. Lett. {\bf 93} (2004) 262002; 
Nucl.\ Phys.\  B {\bf 724} (2005) 197;\\
 C.~Anastasiou, G.~Dissertori and F.~St\"ockli,
  arXiv:0707.2373 [hep-ph].

\bibitem{babismu}
 C.~Anastasiou, K.~Melnikov and F.~Petriello,
  hep-ph/0505069.

\bibitem{babisdy}
 K.~Melnikov and F.~Petriello,
  Phys.\ Rev.\ Lett.\  {\bf 96} (2006) 231803;
  Phys.\ Rev.\  D {\bf 74} (2006) 114017.



\bibitem{chet}
F.V.\ Tkachov, Phys.\ Lett.\ {\bf 100B} (1981) 65;\\
K.G.\ Chetyrkin and F.V.\ Tkachov, Nucl.\ Phys.\ {\bf B192} (1981) 159.

\bibitem{gr} 
T.\ Gehrmann and E.\ Remiddi, Nucl.\ Phys.\ B
{\bf 580} (2000) 485.




\bibitem{laporta}
S.~Laporta,
Int.\ J.\ Mod.\ Phys.\ A {\bf 15} (2000) 5087.


\bibitem{air}
 C.~Anastasiou and A.~Lazopoulos,
  JHEP {\bf 0407} (2004) 046.


\bibitem{smirnovbook}
V.A.\ Smirnov, {\it Evaluating Feynman Integrals}, Springer Tracts of 
Modern Physics (Heidelberg, 2004).



\bibitem{ambre}
J.~Gluza, K.~Kajda and T.~Riemann,
  arXiv:0704.2423 [hep-ph].



\bibitem{daleo}
 C.~Anastasiou and A.~Daleo,
  JHEP {\bf 0610} (2006) 031.

\bibitem{mb}
 M.~Czakon,
  Comput.\ Phys.\ Commun.\  {\bf 175} (2006) 559.



\bibitem{nested}
S.~Moch, P.~Uwer and S.~Weinzierl,
J.\ Math.\ Phys.\  {\bf 43} (2002) 3363;\\
S.~Weinzierl, Comput.\ Phys.\ Commun.\  {\bf 145} (2002) 357.
\bibitem{xsummer}
  S.~Moch and P.~Uwer,
  Comput.\ Phys.\ Commun.\  {\bf 174} (2006) 759.



\bibitem{blumlein}
 J.~Bl\"umlein,
  Comput.\ Phys.\ Commun.\  {\bf 159} (2004) 19.



\bibitem{smirbox}
  V.~A.~Smirnov,
  Phys.\ Lett.\  B {\bf 460} (1999) 397;
  {\bf 491} (2000) 130;
  {\bf 500} (2001) 330;
  {\bf 524} (2002) 129;
  {\bf 567} (2003) 193;\\
  V.A.~Smirnov and O.L.~Veretin,
  Nucl.\ Phys.\  B {\bf 566} (2000) 469;\\
  G.~Heinrich and V.A.~Smirnov,
  Phys.\ Lett.\  B {\bf 598} (2004) 55.

\bibitem{tausk}
J.B.\ Tausk, Phys.\ Lett.\ B {\bf 469} (1999) 225.


\bibitem{oleari2}
C.\ Anastasiou, J.B.\ Tausk and M.E.\ Tejeda-Yeomans,
Nucl.\ Phys.\  B (Proc.\ Suppl.) {\bf 89} (2000) 262.

\bibitem{gluza}
  M.~Czakon, J.~Gluza and T.~Riemann,
  Phys.\ Rev.\  D {\bf 71} (2005) 073009;
  Nucl.\ Phys.\  B {\bf 751} (2006) 1;\\
 S.~Actis, M.~Czakon, J.~Gluza and T.~Riemann,
  arXiv:0704.2400 [hep-ph].

\bibitem{kotikov}
  A.V.~Kotikov,
  Phys.\ Lett.\  B {\bf 254} (1991) 158.

\bibitem{remiddi}
E.~Remiddi,
  Nuovo Cim.\  A {\bf 110} (1997) 1435;\\
M.~Caffo, H.~Czyz, S.~Laporta and E.~Remiddi,
  Nuovo Cim.\  A {\bf 111} (1998) 365.


\bibitem{hypexp}
 T.~Huber and D.~Ma\^{\i}tre,
  Comput.\ Phys.\ Commun.\  {\bf 175} (2006) 122.


\bibitem{mastrolia}
 M.~Argeri and P.~Mastrolia,
  arXiv:0707.4037 [hep-ph].


\bibitem{oleari1}
C.\ Anastasiou, T.\ Gehrmann, C.\ Oleari, E.\ Remiddi and
J.B.\ Tausk, Nucl.\ Phys.\ B
{\bf 580} (2000) 577.



\bibitem{bastei}
T.~Gehrmann and E.~Remiddi, Nucl.\ Phys.\ B (Proc.\ Suppl.)
{\bf 89} (2000) 251.


\bibitem{3jmi}
T.\ Gehrmann and E.\ Remiddi, Nucl.~Phys.~B {\bf 601} (2001) 248;
{\bf 601} (2001) 287.




\bibitem{bonciani}
  R.~Bonciani, P.~Mastrolia and E.~Remiddi,
  Nucl.\ Phys.\  B {\bf 661} (2003) 289; {\bf 702} (2004) 359 (E);
  Nucl.\ Phys.\  B {\bf 690} (2004) 138.


\bibitem{kunszt}
 C.~Anastasiou, S.~Beerli, S.~Bucherer, A.~Daleo and Z.~Kunszt,
  JHEP {\bf 0701} (2007) 082;\\
  U.~Aglietti, R.~Bonciani, G.~Degrassi and A.~Vicini,
  JHEP {\bf 0701} (2007) 021.



\bibitem{bhabhade}
  R.~Bonciani, A.~Ferroglia, P.~Mastrolia, E.~Remiddi and J.J.~van der Bij,
  Nucl.\ Phys.\  B {\bf 701} (2004) 121.




\bibitem{m1}
C.\ Anastasiou, E.W.N.~Glover, C.\ Oleari and M.E.\ Tejeda-Yeomans,
Nucl.\ Phys.\ B~{\bf 601}~(2001) 318;~{\bf 601}~(2001)~347;
 {\bf 605} (2001) 486;\\
E.W.N.~Glover, C.~Oleari and M.E.~Tejeda-Yeomans,
Nucl.\ Phys.\ {\bf 605} (2001) 467;\\
C.~Anastasiou, E.W.N.~Glover and M.E.~Tejeda-Yeomans,
Nucl.\ Phys.\ B {\bf 629} (2002) 255;\\
E.W.N.~Glover and M.E.~Tejeda-Yeomans,
JHEP {\bf 0306} (2003) 033;\\
E.W.N.~Glover,
JHEP {\bf 0404} (2004) 021;\\
Z.~Bern, A.~De Freitas and L.J.~Dixon,
JHEP {\bf 0203} (2002) 018;
JHEP {\bf 0306} (2003) 028;\\
A.~De Freitas and Z.~Bern,
JHEP {\bf 0409} (2004) 039.

\bibitem{m2}
Z.~Bern, A.~De Freitas and L.J.~Dixon,
JHEP {\bf 0109} (2001) 037.

\bibitem{m3}
Z.~Bern, A.~De Freitas, L.J.~Dixon, A.~Ghinculov and H.L.~Wong,
JHEP {\bf 0111} (2001) 031;\\
T.~Binoth, E.W.N.~Glover, P.~Marquard and J.J.~van der Bij,
JHEP {\bf 0205} (2002) 060.

\bibitem{3jme}
L.W.~Garland, T.~Gehrmann, E.W.N.~Glover, A.~Koukoutsakis and E.~Remiddi,
Nucl.\ Phys.\ B {\bf 627} (2002) 107;
{\bf 642} (2002) 227.


\bibitem{muw}
S.~Moch, P.~Uwer and S.~Weinzierl,
Phys.\ Rev.\ D {\bf 66} (2002) 114001.



\bibitem{ancont}
T.~Gehrmann and E.~Remiddi,
Nucl.\ Phys.\ B {\bf 640} (2002) 379.


\bibitem{breuther}
W.~Bernreuther, R.~Bonciani, T.~Gehrmann, R.~Heinesch, T.~Leineweber, P.~Mastrolia and E.~Remiddi,
  Nucl.\ Phys.\  B {\bf 706} (2005) 245;
 {\bf 712} (2005) 229;
  Phys.\ Rev.\ Lett.\  {\bf 95} (2005) 261802;
  Nucl.\ Phys.\  B {\bf 750} (2006) 83;\\
 W.~Bernreuther, R.~Bonciani, T.~Gehrmann, R.~Heinesch, T.~Leineweber and E.~Remiddi,
  Nucl.\ Phys.\  B {\bf 723} (2005) 91;\\
  W.~Bernreuther, R.~Bonciani, T.~Gehrmann, R.~Heinesch, P.~Mastrolia and E.~Remiddi,
  Phys.\ Rev.\  D {\bf 72} (2005) 096002.



\bibitem{bhabha}
  R.~Bonciani and A.~Ferroglia,
  Phys.\ Rev.\  D {\bf 72} (2005) 056004.




\bibitem{hpl}
E.\ Remiddi and J.A.M.\ Vermaseren, Int.\ J.\ Mod.\ Phys.\ A {\bf 15}
(2000) 725;\\
T.~Gehrmann and E.~Remiddi,
Comput.\ Phys.\ Commun.\ {\bf 141} (2001) 296;\\
D.\ Ma\^{\i}tre, 
  Comput.\ Phys.\ Commun.\  {\bf 174} (2006) 222;
  hep-ph/0703052.


\bibitem{nielsen}
N.~Nielsen, Nova Acta Leopoldiana (Halle) {\bf 90} (1909) 123;\\
K.S.\ K\"olbig, J.A.\ Mignaco and E.\ Remiddi, BIT {\bf 10} (1970) 38.


\bibitem{tdhpl}
T.~Gehrmann and E.~Remiddi,
  Comput.\ Phys.\ Commun.\  {\bf 144} (2002) 200.


\bibitem{weinzierl}
 J.~Vollinga and S.~Weinzierl,
  Comput.\ Phys.\ Commun.\  {\bf 167} (2005) 177.




\bibitem{catani}
S.\ Catani, Phys.\ Lett.\ B {\bf 427} (1998) 161;\\
G.~Sterman and M.E.~Tejeda-Yeomans,
Phys.\ Lett.\ B {\bf 552} (2003) 48.

\bibitem{mm}
 A.~Mitov and S.~Moch,
  JHEP {\bf 0705} (2007) 001.

\bibitem{top}
 M.~Czakon, A.~Mitov and S.~Moch,
  Phys.\ Lett.\  B {\bf 651} (2007) 147;
arXiv:0707.4139 [hep-ph].

\bibitem{penin}
 A.A.~Penin,
  Phys.\ Rev.\ Lett.\  {\bf 95} (2005) 010408;
  Nucl.\ Phys.\  B {\bf 734} (2006) 185.

\bibitem{becher}
  T.~Becher and K.~Melnikov,
  JHEP {\bf 0706} (2007) 084.



\bibitem{sub}
Z.~Kunszt and D.E.~Soper,
Phys.\ Rev.\ D {\bf 46} (1992) 192.


\bibitem{cs}
S.~Catani and M.H.~Seymour,
Nucl.\ Phys.\ B {\bf 485} (1997) 291; {\bf 510} (1997) 503(E).

\bibitem{singleun}
S.~Frixione, Z.~Kunszt and A.~Signer,
Nucl.\ Phys.\ B {\bf 467}, 399 (1996);\\
G.~Somogyi and Z.~Trocsanyi,
Acta Phys.\ Chim.\ Debr.\ {\bf XL} (2006) 101;\\
 Z.~Nagy, G.~Somogyi and Z.~Trocsanyi,
  hep-ph/0702273.

\bibitem{ant}
D.A.~Kosower,
Phys.\ Rev.\ D {\bf 57} (1998) 5410;
{\bf 71} (2005) 045016.

\bibitem{hadant}
 A.~Daleo, T.~Gehrmann and D.~Ma\^{\i}tre,
  JHEP {\bf 0704} (2007) 016.


\bibitem{nnlosub}
S.~Weinzierl,
JHEP {\bf 0303} (2003) 062;\\
D.A.~Kosower, Phys.\ Rev.\ D {\bf 67} (2003) 116003;\\
W.B.~Kilgore,
Phys.\ Rev.\ D {\bf 70} (2004) 031501;\\
M.\ Grazzini and S.\ Frixione,
JHEP {\bf 0506} (2005) 010;\\
G.~Somogyi, Z.~Trocsanyi and V.~Del Duca,
JHEP {\bf 0506} (2005) 024;  {\bf 0701} (2007) 070;\\
G.~Somogyi and Z.~Trocsanyi,
JHEP {\bf 0701} (2007) 052.





\bibitem{cghiggs}
 S.~Catani and M.~Grazzini,
  Phys.\ Rev.\ Lett.\  {\bf 98} (2007) 222002.


\bibitem{ourant}
A.~Gehrmann-De Ridder, T.~Gehrmann and E.W.N.~Glover,
  JHEP {\bf 0509} (2005) 056.



\bibitem{our2j}
A.~Gehrmann-De Ridder, T.\ Gehrmann and E.W.N.\ Glover, 
Nucl.\ Phys.\ B {\bf 691} (2004) 195;
Phys.\ Lett.\ B {\bf 612} (2005) 36;
{\bf 612} (2005) 49.



\bibitem{ourT}
A.~Gehrmann-De Ridder, T.~Gehrmann, E.W.N.\ Glover and G.~Heinrich,
  arXiv:0707.1285 [hep-ph] and these proceedings.


\bibitem{weinzierl2j}
S.~Weinzierl,
  Phys.\ Rev.\ D {\bf 74} (2006) 014020.

\bibitem{our3j}
A.~Gehrmann-De Ridder, T.~Gehrmann and E.W.N.~Glover,
Nucl.\ Phys.\ Proc.\ Suppl.\  {\bf 135} (2004) 97;\\
A.~Gehrmann-De Ridder, T.~Gehrmann, E.W.N.~Glover and G.~Heinrich,
  Nucl.\ Phys.\ Proc.\ Suppl.\  {\bf 160} (2006) 190.




\bibitem{afbnnlo}
 S.~Catani and M.H.~Seymour,
  JHEP {\bf 9907} (1999) 023;\\
 S.~Weinzierl,
  Phys.\ Lett.\  B {\bf 644} (2007) 331.





\end{thebibliography}
\end{document}